\documentclass[aps,prl,twocolumn,showpacs]{revtex4-1}

\pdfoutput=1

\usepackage{graphicx,color}			
\usepackage{amsmath}			
\usepackage{amsfonts}
\usepackage{amssymb}
\usepackage{amsthm}
\usepackage{verbatim}

\newcommand{\bqa}{\begin{eqnarray}}
\newcommand{\eqa}{\end{eqnarray}}
\newcommand{\nn}{\nonumber \\}
\newcommand{\beq}{\begin{equation}}
\newcommand{\eeq}{\end{equation}}

\renewcommand{\vec}[1]{{\boldsymbol{#1}}}

\def\be{\begin{eqnarray}}
\def\ee{\end{eqnarray}}

\usepackage{color}
\definecolor{darkblue}{rgb}{0.,0.,0.4}
\definecolor{darkred}{rgb}{0.5,0.,0.}
\definecolor{BlueViolet}{RGB}{138,43,226}
\definecolor{SkyBlue}{RGB}{30,144,255}
\definecolor{DarkGreen}{RGB}{0,100,0}
\usepackage[pdftex,colorlinks=true,linkcolor=darkblue,citecolor=blue,urlcolor=darkred]{hyperref}

\begin{document}

\title{Exceptional points for chiral Majorana fermions in arbitrary dimensions}

\author{Ipsita Mandal}
\affiliation{Perimeter Institute for Theoretical Physics, 31 Caroline St. N., Waterloo ON N2L 2Y5, Canada}
\email{imandal@pitp.ca}

\date{\today}

\begin{abstract}
Certain real parameters of a Hamiltonian, when continued to complex values, can give rise to singular points called exceptional points ($EP$'s), where two or more eigenvalues coincide and the complexified Hamiltonian becomes non-diagonalizable. We show that for a generic $d$-dimensional topological superconductor/superfluid with a chiral symmetry, one can find $EP$'s associated with the chiral zero energy Majorana fermions bound to a topological defect/edge. Exploiting the chiral symmetry, we propose a formula for counting the number ($n$) of such chiral zero modes. We also establish the connection of these solutions to the Majorana fermion wavefunctions in the position space. The imaginary parts of these momenta are related to the exponential decay of the wavefunctions localized at the defect/edge, and hence their changes of signs at a topological phase transition point signal the appearance or disappearance of chiral Majorana zero modes. Our analysis thus explains why topological invariants like the winding number, defined for the corresponding Hamiltonian in the momentum space for a defectless system with periodic boundary conditions, capture the number of admissible Majorana fermion solutions for the position space Hamiltonian with defect(s). Finally, we conclude that $EP$'s cannot be associated with the Majorana fermion wavefunctions for systems with no chiral symmetry, although one can use our formula for counting $n$, using complex $k$ solutions where the determinant of the corresponding BdG Hamiltonian vanishes.
\end{abstract}


\maketitle

\paragraph{\textbf{Introduction --}}
The Hamiltonian operator can contain certain real parameters, which on being continued to complex values, give rise to singular points where the operator becomes non-diagonalizable. These are called exceptional points ($EP$'s), at which two or more of the eigenvalues coalesce and the norm of at least one eigenvector of the complexified Hamiltonian vanishes \cite{Panch,Berry,Kato,Rotter,Heiss,Fagotti,Brand}.
The concept of $ EP $'s is similar to that of a degeneracy point,
but with the important difference that all the energy eigenvectors cannot be made mutually orthogonal. In previous works, $EP$'s have been used \cite{sourin,ramon,dmitry1,dmitry2,ipsita} to
describe topological phases of matter associated with zero energy Majorana bound
states (MBSs) in one-dimensional ($1d$) topological superconductors/superfluids.

The first-quantized Hamitonians describing fully gapped noninteracting topological insulators and superconductors in $d$-dimensions can be classified into ten symmetry classes \cite{Schnyder_2008,Ryu_2010} in terms of nonspatial symmetries, i.e., symmetries that act locally in the position space, namely time-reversal symmetry (TRS), particle-hole symmetry (PHS), and chiral symmetry. Recently, it has been realized \cite{chiu2013,chiu2014,morimoto2013,kane2013,kane-mirror,bena,mirror2015} that the complete classification should include topological states protected by crystalline symmetries (such as mirror reflections and rotations), which are spatial symmetries acting nonlocally in the position space.

A superconductor is described by a Bogoliubov de Gennes (BdG) Hamiltonian ($H_{\mbox{\tiny{BdG}}}$), which has an exact PHS, in addition to having the structure of an ordinary Bloch Hamiltonian. The PHS of $H_{\mbox{\tiny{BdG}}}$ implies
\begin{eqnarray}
\mathcal{C} \, H_{\mbox{\tiny{BdG}}}( \vec k) \, \mathcal{C}^{-1} &=& - H_{\mbox{\tiny{BdG}}}^* (- \vec k) \,, \,
\mathcal{C} = \left(
\begin{array}{cc}
 0 & \mathbb{I}  \\
 \mathbb{I}   &  0   \\
\end{array} \right) ,
\end{eqnarray}
where $\mathcal{C}$ is an antiunitary operator, so that the energy eigenvalues $\pm E(\vec k)$ always appear in pairs for each $\vec k$-value. In this paper, we will consider the case $\mathcal{C}^2 = 1$ for which $H_{\mbox{\tiny{BdG}}}$ can be categorized into three non-trivial classes: BDI, DIII and D.

Topologically protected gapless modes can occur at a topological defect or surface of a topological superconductor/insulator. 
%A classification of topological defects and their associated gapless modes for $d$-dimensional Bloch-BdG  Hamiltonians $H(\vec k, \vec r)$ is known \cite{teo-kane}, where $\vec k$ is the momentum in a $d$-dimensional Brillouin zone (defined on a torus $T^d$), and $\vec r$ characterizes the $D$-dimensional surface $S^D $ surrounding the defect. In this terminology, a point corresponds to $d - D = 1$, a line defect has $d - D = 2$, while a planar defect is defined by $d-D=3$. 
%This classification helps in establishing the bulk-boundary correspondence relating the topological class of the bulk Hamiltonian to the structure of the protected modes associated with the defect. Recently, more examples of topologically protected zero mode solutions \cite{fukui} have been studied, which were not included in the classification table found by Teo and Kane.
Due to the PHS, the boundary excitations of a topological superconductor are Majorana fermions, such that the creation operator ($\gamma^{\dagger}_{E=0}$) is equivalent to the annihilation operator ($\gamma_{E=0}$). Since they obey non-Abelian statistics, they might have potential applications in designing fault-tolerant quantum computers \cite{Kitaev_2001,Nayak_2008}.

%Recently, it has been realized \cite{chiu2013,morimoto2013,kane2013,chiu2014,kane-mirror,bena,mirror2015} that the complete classification should include topological states protected by crystalline symmetries (such as mirror reflections and rotations), which are spatial symmetries acting nonlocally in the position space. Let us consider a $(p-1)$-dimensional plane embedded in $p$ spatial dimensions and invariant under a reflection of one of the spatial directions. For any defect lying on a $(p-1)$-dimensional plane orthogonal to this reflection-symmetric plane, one can find a gapless boundary mode protected by the reflection, just like for the case of topological phases protected by nonspatial discrete symmetries. They are characterized by an integer ($\mathbb{Z}$) or $\mathbb{Z}_2$ ``mirror" topological invariant. Such zero modes at the edges of $1d$ and $2d$ Rashba semiconductor deposited on iron-based superconductor have been studied \cite{kane2013,bena,mirror2015}.

In an earlier work \cite{ipsita}, a formula to count the Majorana zero modes in $1d$ chiral topological superconductors/superfluids based on the notion of $EP$'s has been proposed. This formula serves as an alternative to the familiar winding number ($\mathbb{W}$) \cite{Schnyder_2008,Ryu_2010,TewariPRL2012}. In this paper, we will show that we can apply the $EP$ formalism to the zero energy Majorana excitations bound to defects for generic $d$-dimensional Hamiltonians anticommuting with a chiral symmetry operator. We will also generalize the formula to count the number ($n$) of Majorana fermions in the presence or absence of the chiral symmetry.

%The paper is organized as follows: In Sec.~\ref{EPsoln}, we show the connection between $EP$ formalism, for a chiral $H_{\mbox{\tiny{BdG}}}$, to the Majorana fermion solutions bound to a topological defect, associated with a $d$-dimensional topological superconductor. In Sec.~\ref{epnonchiral}, we study systems without any chiral symmetry and prove that $EP$'s cannot be associated with Majorana zero modes. In Sec.~\ref{point}, we consider point defects for various classes of $H_{\mbox{\tiny{BdG}}}$. Sec.~\ref{line} is devoted to the study of line defects. We conclude with a summary and outlook in Sec.~\ref{conclusion}.

%%%%%%%%%%%%%%
\paragraph{\textbf{EP solutions for systems with a chiral symmetry --}}

We consider a topological defect embedded in (or at the boundary of) a $d$-dimensional topological superconductor. Let $m$ be the dimensions of the defect, parametrized by the Cartesian coordinates $\vec {r}_{\perp} =\left (  r_{1}, \ldots, r_{d-m}   \right )$ and $\vec {r}_{\parallel} =\left (  r_{d-m+1}, \ldots, r_d   \right )$, located at $\vec r_{\perp} =\vec 0 $. Let $\vec {k}_{\perp} =k_{\perp} \hat {\vec \Omega} = \left (  k_{1}, \ldots, k_{d-m}   \right ) $ and $\vec {k }_{\parallel} =\left (  k_{d-m+1}, \ldots, k_d   \right )$ be the corresponding conjugate momenta, where $k_{\perp} = |\vec k_{\perp}|$ and $ \hat {\vec \Omega}$ is the unit vector when written in spherical coordinates. If there is a chiral symmetry operator $\mathcal{O}$ which anticommutes with the Hamiltonian, the latter takes the form 
\begin{equation}
H_{\mbox{\tiny{chiral}}} = \left(
\begin{array}{cc}
 0 & \mathcal{A} (-i \vec \nabla) \\
 \mathcal{A}^{\dagger} (-i \vec \nabla)   &  0   \\
\end{array} \right) ,
\label{chiralbasis}
\end{equation}
%%%%%%%%%%%%%%%%
in the eigenbasis of $\mathcal{O}$ (where it is diagonal). Here $\mathcal{A} $ is an $ N \times N$ matrix operator in the position space. The solutions for the Majorana zero modes are given by $\psi_{+} =(u_+,0)^T$ and $\psi_{-} =(0,u_-)^T$ (with chirality $+1$ and $-1$ respectively, since $\mathcal O \, \psi_{\pm} = \pm \, \psi_{\pm} $, satisfying:
\begin{eqnarray}
\mathcal{A} (-i \vec \nabla) \, u_{-} =0 \,,
\quad \mathcal{A}^{\dagger} (-i \vec \nabla)  \, u_{+} =0 \,.
\end{eqnarray}
Assuming the dependence $u_- \sim \exp \left( - z^{\alpha} r^{\alpha} \right)$ in the bulk, we find that $z^\alpha$'s should satisfy
\begin{eqnarray}
&& \det \left( \mathcal{A} (i \vec{z}) \right) =0 \,.
\label{deta}
\end{eqnarray}

Now let us find out the $EP$ solutions for the Hamiltonian 
\begin{equation}
H_{\mbox{\tiny{chiral}}} (\vec k) = \left(
\begin{array}{cc}
 0 & \mathcal{A}  (\vec k) \\
 \mathcal{A}^{\dagger} (\vec k)  &  0   \\
\end{array} \right) ,
%\quad \mathcal{A}^{\dagger} (\vec k) = \mathcal{A}^T (-\vec k)\,,
\label{chiralmom}
\end{equation}
in the momentum space, for the corresponding system with no defect. On analytically continuing the magnitude $k_{\perp} \equiv k= |\vec k| $ to the complex $k_{\perp}$-plane, at least one of the eigenvectors of $H_{\mbox{\tiny{chiral}}}( \vec k)$ collapses to zero norm where
\begin{equation} 
\det \left( \mathcal{A} (\vec{k}) \right) =0 \, , \,\, \mbox{or} \,
\det \left(  \mathcal{A}^{\dagger} (\vec{k}) \right) =0 \,.
\label{detak}
\end{equation} 
%%%%%%%%%%%%%%%%%%%%%%%
These points are associated with the solutions of $EP$'s for complex $k_{\perp}$-values where two or more energy levels coalesce. Furthermore, these coalescing eigenvalues have zero magnitude as $\det \left( \mathcal{A} ( \vec{k} ) \right) =0$ (or $ \det \left( \mathcal{A}^{\dagger} (\vec{k})\right) =0$) also implies $\det \left( H_{\mbox{\tiny{chiral}}} (\vec{k}) \right) =0 $. $H_{\mbox{\tiny{chiral}}} (\vec k)$ becomes non-diagonalizable, as in the complex $k_{\perp}$-plane, $\det \left( \mathcal{A} (\vec{k}) \right) = 0 \nRightarrow  \det \left( \mathcal{A}^{\dagger} (\vec{k}) \right) = 0$  (or vice versa). However, at the physical phase transition points, the imaginary parts of one or more solutions vanish, and $\det \left( \mathcal{A} ( \vec{k} ) \right) = \det \left( \mathcal{A}^{\dagger} (\vec{k}) \right) = 0$ for those solutions, making $H_{\mbox{\tiny{chiral}}} (\vec k)$ once again diagonalizable and marking the disappearance of the corresponding $EP$'s. We find that both $ k_{\perp}$  and $i \, z$ (where $z  = |\vec z|$) are obtained as the solutions of the same equation (Eq.~(\ref{deta})), immediately indicating the correspondence $ i \, k_{\perp} \leftrightarrow - z$. Each $EP$ solution corresponds to a Majorana fermion of a definite chirality.

For a generic $H_{\mbox{\tiny{chiral}}}$, let $ k_{A}^{j} $ and $k_{B }^{j}$  ($j = 1, \ldots , Q $) be the two sets of $EP$ solutions for $\det \left( \mathcal{A} (\vec{k}) \right) =0 $ and $\det \left(  \mathcal{A}^{\dagger} (\vec{k}) \right) =0$ respectively, related by $ \lbrace \operatorname{Im} (k_{ A }^{j} ) \rbrace = - \lbrace  \operatorname{Im} ( k_{ B }^{j } ) \rbrace$, after $k_{\perp}$ has been analytically continued to the complex plane. At a topological phase transition point, one or more of the $\operatorname{Im} \left(  k_{A/B}^j \right) $'s go through zero. When $\operatorname{Im} \left(  k_{A/B}^j \right) $ changes sign at a topological phase transition point, the position space wavefunction of the corresponding Majorana fermion changes from exponentially decaying to exponentially diverging or vice versa. If the former happens, the Majorana fermion ceases to exist. A new Majorana zero mode appears in the latter case. The count ($n$) for the Majorana fermions for a defect is thus captured by the function
\begin{eqnarray}
\label{fchiral1}
f(\lbrace \lambda_i \rbrace, \vec k_{\parallel} , \hat{\vec{\Omega}} ) &=&  \frac{1} {2} \,
\Big | \sum_{j=1}^{Q} \Big[\, sgn \big \lbrace \operatorname{Im} \left ( k_{A/B}^{j} \left( \lbrace \lambda_i \rbrace   , \vec k_{\parallel} , \hat{\vec{\Omega}} \right) \right ) \big \rbrace \nn 
&& \quad \quad - sgn \big \lbrace \operatorname{Im} \left ( k_{A/B}^{j} \left ( \lbrace \lambda_i^0 \rbrace   , \vec k_{\parallel}^0 , \hat{\vec{\Omega}}^0 \right ) \right ) \big \rbrace  
\Big] \,  \Big |\,,\nn
\end{eqnarray}
%%%%%%%%%%%%%%%%%%%%%%%%
where $(\lbrace \lambda_i \rbrace, \vec k_{\parallel} , \hat{\vec{\Omega}} ) $ are the parameters appearing in the expressions for $k_{A/B}^{j}$, and $ \left ( \lbrace \lambda_i^0 \rbrace   , \vec k_{\parallel}^0 , \hat{\vec{\Omega}}^0  \right) $ are their values at any point in the non-topological phase. If $\mathcal{A}^{\dagger} (\vec{k})= \mathcal{A}^{T} (\vec{-k})$ holds, then the two sets of $EP$'s are related by $ \lbrace k_{ A }^{j} \rbrace = - \lbrace  k_{ B }^{j } \rbrace$, one set corresponding to the the solutions obtained from one of the two off-diagonal blocks. In such cases, the pairs of the Majorana fermion wavefunctions are of opposite chiralities.

This connection between the complex $k_{\perp}$-space solutions where two or more zero energy eigenvalues coalesce, with the Majorana fermion solutions of the position space Hamiltonian with a topological defect, helps us understand why the topological invariants like $\mathbb{W}$ (winding number \cite{TewariPRL2012}) and $f(\lbrace \lambda_i\rbrace  , \vec k_{\parallel} , \hat{\vec{\Omega}}  )$, defined for systems with a chiral symmetry, are related to $n$. $\mathbb{W}$ and $f(\lbrace \lambda_i \rbrace, \vec k_{\parallel} , \hat{\vec{\Omega}} )$ are defined in terms of $\det \left( \mathcal{A} ( \vec{k} ) \right) $ for $H_{\mbox{\tiny{chiral}}} $ written in the momentum space, in a given topological phase. On the other hand, Majorana fermion solutions are calculated for the corresponding Hamiltonian at the location of the defect. Along an $EP$ solution, the complexified Hamiltonian has vanishing determinant, indicating the presence of two or more coalescing zero eigenvalues. The actual physical Hamiltonian, however, has zero determinant only for a real $k_{\perp}$-value, at which one or more $EP$'s collapse, allowing the Hamiltonian to be diagonalizable at that point. Observing the correspondence $\exp \left( i\, k_{A/B}^j \right)  \Leftrightarrow \exp \left(  - z_j \, r_{\perp} \right) $, such that $\sum_j a_j \, \exp \left(  - z_j \, r_{\perp}  \right) $ is a Majorana fermion solution, explains why a Majorana zero energy state exists throughout a topological phase, with the topological invariants $f( \lbrace \lambda_i\rbrace  , \vec k_{\parallel} , \hat{\vec{\Omega}})$ and $\mathbb{W}$ capturing the number of such modes, and changing their values only on crossing a phase transition point. This also provides an intuitive understanding of the bulk-edge correspondence.

%%%%%%%%%%%%%%%%%%
\paragraph{\textbf{EP solutions for systems without chiral symmetry --}}
\label{epnonchiral}

A Hamiltonian $H_D$ for a system without any chiral symmetry can be constructed from $H_{\mbox{\tiny{chiral}}}$ as \cite{teo-kane}
\begin{eqnarray}
\label{nonchiral}
H_D &=& \cos \theta \, H_{\mbox{\tiny{chiral}}} + \sin \theta \, \mathcal{O} \,,\nn
&=& \cos \theta \,\left(
\begin{array}{cc}
 0 & \mathcal{A}  \\
 \mathcal{A}^{\dagger}   &  0   \\
\end{array} \right) 
 + \sin \theta \, 
 \left(
\begin{array}{cc}
 \mathbb{I} & 0  \\
 0  &   - \mathbb{I}   \\
\end{array} \right) \,.
\end{eqnarray}
This clearly breaks the chiral symmetry unless $\theta = 0$. The $EP$'s are given by $\det \left( \mathcal{A}(\vec k) \right )=0$ or $\det \left( \mathcal{A}^{\dagger}(\vec k) \right )=0$, where two levels coalesce for a complex value of $k_{\perp}$. However, we immediately observe that these $EP$'s do not correspond to zero energy modes for the complexified $H_D (\vec k)$. In the momentum space, at the points of vanishing energy eigenvalues (and hence vanishing $\det \left( H_D (\vec k) \right)  $), $H_D (\vec k) $ still remains perfectly diagonalizable even for complex $k_{\perp}$-values.

The position space Majorana fermion solutions thus cannot be captured by the $EP$ solutions in the complex $ k_{\perp}$-space, as the latter are related to vanishing eigenvectors for the complexified Hamiltonian where it becomes non-diagonalizable. However, we can still encode the count of the non-chiral Majorana zero modes by the function defined in Eq.~(\ref{fchiral1}), but with $ k_{A/B}^{j} $ being the two sets of complex-valued solutions for $\det \left( H_D (\vec k) \right) = 0 $, obeying $ \lbrace \operatorname{Im} (k_{ A }^{j}) \rbrace = - \lbrace \operatorname{Im} ( k_{ B }^{j } ) \rbrace$.

%%%%%%%%%%%%%%
\paragraph{\textbf{Point Defects --}}
\label{point}

In this section, we consider point defects for a $d$-dimensional BdG Hamiltonian with the $PHS$ operator $\mathcal{C}$ squaring to $+1$. Such a defect can appear at the edges of a $1d$ chain, as vortex solutions in a $2d$ system, or as hedgehog configurations in a $3d$ bulk. There can be a Majorana zero energy state bound to such a topological defect, depending on the values of the parameters. 

%%%%%%%%%%%%%55
\paragraph{Point Defects in class BDI --}
\label{pointb}

The BDI symmetry class is associated with the existence of topologically protected chiral Majorana fermions and characterized by an integer ($\mathbb{Z}$) topological invariant.

We consider a specific model \cite{fujiwara} for such point defects described by a BdG-Dirac type
Hamiltonian:
\begin{equation}
H_{\mbox{\tiny{chiral}}} = -i \Gamma^\alpha \, \partial_\alpha + \Gamma^{d+\beta} \phi_\beta \,,
\label{dirac}
\end{equation}
where $\alpha, \beta  =1, \dots , d$, $\phi_\beta$ is a component of the generalized $d$-component order parameter $\vec \phi$, and the $2d$-dimensional gamma matrices obey the anticommutation relations $\lbrace \Gamma^\mu , \Gamma^\nu \rbrace = 2 \, \delta^{\mu \nu}$. Clearly, $ \Gamma^\alpha$ and $ \Gamma^\beta$ are the sets of gamma matrices associated with the kinetic term and the order parameter part of $H_{\mbox{\tiny{chiral}}}$.

The point defect can be modelled by choosing $\vec \phi$ of the form:
\begin{eqnarray}
\phi_\alpha ( \vec r) = \Delta ( r ) \, \hat{r}^\alpha \,, \quad
\Delta (r) = \begin{cases} 
0 & \text{for }   r< R \,, \\
\Delta_0 & \text{for }   r \geq  R \,,
  \end{cases} 
\end{eqnarray}
%%%%%%%%%%%%%%%%%%%%%%
where $\Delta_0$ is a constant.
The chirality operator $\mathcal{O}$ can be identified with
\begin{equation}
\Gamma^{2d + 1} = (-i)^d \, \Gamma^1 \cdots \Gamma^{2d} \,,
\end{equation}
which anticommutes with all the $\Gamma^\mu$'s.

One can choose the gamma matrices as follows
\footnote{\unexpanded{Let us define the $\gamma$-matrices for $d=2 \ell$ (even) as follows:
\begin{eqnarray}
\gamma^{2j -1 } &=& \underbrace{\mathbb{I} \otimes \cdots \mathbb{I}}_{j-1}
\otimes  \sigma^1 \otimes 
\underbrace{\sigma^3 \otimes \cdots \otimes \sigma^3}_{ \ell-j} \,,\nn
%%%%%%%%%%
\gamma^{2j } &=& \underbrace{\mathbb{I} \otimes \cdots \mathbb{I}}_{j-1}
\otimes  \sigma^2 \otimes 
\underbrace{\sigma^3 \otimes \cdots \otimes \sigma^3}_{  \ell-j} \,,\nn
\end{eqnarray}
%%%%%%%%%%%%%%%%%%%%
where $j=1, \ldots,  \ell $.
If $d=2 \ell+1$ (odd), then we need to add another $\gamma$-matrix
\begin{eqnarray}
\gamma^{2 \ell+1 } &=& \underbrace{\sigma^3 \otimes \cdots \sigma^3}_{ \ell}
=(- i )^n \gamma^1 \cdots \gamma^{2 \ell} \,.
\end{eqnarray}
%%%%%%%%%%%%%%%
Defining
\begin{eqnarray}
\Gamma = \gamma^1 \gamma^3 \cdots \gamma^{2 \ell+1} \,,
\label{Gamma}
\end{eqnarray}
%%%%%%%%%%%%%%
we get
\begin{eqnarray}
\Gamma^2 &=& (-1)^{ \ell \,( \ell+1)/2} \,,\nn
\left( \gamma^{\mu} \right)^T &=& \left( \gamma^{\mu} \right)^{*} =(-1)^ \ell \Gamma \gamma^\mu \Gamma^{-1} \,,
\end{eqnarray}
where $\mu = 1,2 , \ldots , 2 \ell , 2 \ell+1 $.
} }
:\begin{eqnarray}
&& \Gamma^{\alpha} = \gamma^\alpha \otimes \mathbb{I} \otimes \sigma_1 \,, \quad
\Gamma^{d+\alpha} = \mathbb{I} \otimes \gamma^\alpha \otimes \sigma_2 \,, \nn
&& \Gamma^{2d+1} = \mathbb{I} \otimes \mathbb{I} \otimes \sigma_3 \,.
\end{eqnarray}
%where the $\gamma^\alpha$'s have been defined in Appendix~\ref{app:clifford}.

The normalizable solutions \cite{fujiwara} are given by
\begin{eqnarray}
&& \Psi_+ \nn
&=& C_+ \left(
\begin{array}{c}
  \Gamma \\
  0 \\
\end{array} \right) 
 \exp \Big [ (-1)^{ \ell} \int_0^r dr'  \Delta (r')  \Big] \,\, \mbox{for } (-1)^{ \ell}  \Delta_0  < 0 \,, \nn
%%%%%%%%%%
&&\Psi_- \nn
&=& 
C_- \left(
\begin{array}{c}
 0 \\
   \Gamma \\
\end{array} \right)
\exp \Big [ (-1)^{ \ell+1} \int_0^r dr'  \Delta (r')  \Big] 
\, \,\mbox{for } (-1)^{ \ell}  \Delta_0 > 0 \,, \nn
\end{eqnarray}
where $C_{\pm}$ are constants, and $ \ell $ and $\Gamma $ have been defined in Eq.~(\ref{Gamma}). Hence the admissible solutions decay as $\exp (- |\Delta_0|\, r_\perp)$ in the bulk, where $i\, z = i \, \Delta_0$ can be identified with the $EP$ solution with complex $k_\perp$.

In $1d$, numerous lattice models have been studied which can support one \cite{Kitaev_2001} or multiple MBSs \cite{sudip,diptiman1,diptiman2,sumanta-chiral,flensberg,abel} at one edge of an open chain. For such models, one can show that the chiral MBS solutions at an edge, obtained by the transfer matrix approach formalism \cite{Kitaev_2001,sudip,diptiman2011,diptiman1,diptiman2} in the lattice space, have the same relation with the $EP$ solutions obtained in the complex $k$-space.

\paragraph{Point Defects in class D --}
\label{pointd}

In the symmetry class D, the TRS is broken, thus breaking the chiral symmetry, and characterized by a $\mathbb{Z}_2$ topological invariant \cite{teo-kane}, with $n$ allowed to be $0$ or $1$. A topologically protected non-chiral MBS bound to a point defect can exist. In the absence of $\mathcal{O}$, the conclusions for $H_D$ in Eq.~(\ref{nonchiral}) will apply.

%%%%%%%%%%%%%%%%%%%%%%%%%%%%%%%%%

%%%%%%%%%%%%%%%%%%%%%%%%%%%%%%%%
\paragraph{Point defects in class DIII --}
\label{pointd3}
A point defect in class DIII can support a Majorana Kramers pair (MKP) corresponding to doubly degenerate Majorana zero modes, and is characterized by a $\mathbb{Z}_2$ topological invariant. A chiral symmetry operator $\mathcal{O}$ can be defined such that the Hamiltonian in class DIII can be brought to the block off-diagonal form (Eq.~(\ref{chiralbasis})), just like for the class BDI. Furthermore, the DIII class with a mirror symmetry is equivalent to the BDI class with an additional (pseudo) TRS, and hence the edge of a $1d$ Hamiltonian with a mirror line can be characterized by an integer ($\mathbb{Z}$) mirror winding number \cite{kane-mirror,bena,mirror2015} related to the existence of multiple MKPs. Hence the analysis and conclusions for $H_{\text{\tiny{chiral}}}$ in Eq.~(\ref{chiralbasis}) will be applicable for the DIII class. Due the connection between $EP$'s and the Majorana fermion wavefunctions, we can again use Eq.~(\ref{fchiral1}) to count the number of MKPs in a given topological phase in such a $1d$ system. Each Majorana zero mode is of a definite chirality with respect to $\mathcal{O}$.

\paragraph{\textbf{Line Defects --}}
\label{line}
%%%%%%%%%%%%%%%%%

In this section, we consider line defects for a $d$-dimensional BdG Hamiltonians with $\mathcal{C}^2=1$. Majorana zero modes bound to such a defect can appear at the edges of a $2d$ system, or along a vortex line in a $3d$ system \cite{fukui}.

Whenever $H_{\mbox{\tiny{BdG}}}$ cannot be written in a block-off diagonal form in the absence of a chiral symmetry, the system will conform with the discussion for $H_D$ in Eq.~(\ref{nonchiral}).

For a system in class BDI or class DIII, the chiral symmetry operator $\mathcal{O}$ exists, so that the Hamiltonian $H_{\mbox{\tiny{chiral}}}$ takes the block off-diagonal form shown in Eq.~(\ref{chiralbasis}). The allowed solutions will be of the form $\sim \exp \left( i \, k_{\parallel} \, r_{\parallel} \right) \, \exp \left( - z  \, |r_{\perp}| \right )$. Clearly, $  i \,z $ will correspond to the $EP$ solutions in the complex $k_{\perp} $-plane satisfying Eq.(\ref{detak}). Using these solutions, $f( \lbrace \lambda_i\rbrace  ,  k_{\parallel} , \hat{\vec{\Omega}})$ in Eq.~(\ref{fchiral1}) will give the count of the chiral Majorana fermions.

%%%%%%%%%%%%%%%%%%%%%%%%%%%%%

\paragraph{\textbf{Conclusion --}}
\label{conclusion}

We have established the relation of the $EP$ solutions for complexified momenta to the Majorana fermion wavefunctions bound to a topological defect, for a generic $d$-dimensional topological superconductor having a chiral symmetry operator $\mathcal{O}$. This connection explains why topological invariants such as the winding number, defined for the corresponding BdG Hamiltonian without the defect (or edge) written in the momentum space, capture the number of admissible Majorana zero mode solutions for the position space Hamiltonian describing the defect. Each of these solutions are of a definite chirality (with respect to $\mathcal{O}$). We have also proposed a formula for counting the number of Majorana zero modes at each defect/edge, based on these $EP$ solutions. We have shown that such $EP$ solutions cannot exist for systems without any chiral symmetry (e.g.\ class D). Despite such solutions not existing in the absence of $\mathcal{O}$, we have, nevertheless, established an expression for calculating the non-chiral Majorana fermions at a defect from the solutions of $\det \left ( H_{\mbox{\tiny{BdG}}} (\vec k) \right ) = 0 $ in the complexified momentum space. Though we have explicitly discussed only point and line defects, the arguments presented will hold for any $m$-dimensional defect embedded in a $d$-dimensional superconductor (with $d>m$), or for the $(d-1)$-dimensional boundary of the bulk system. A variety of $1d$ and $2d$ models has been studied in \cite{ipsita-example}, explicitly demonstrating how our counting formula works. A proof of the counting formula in Eq.~(\ref{fchiral1}) has also been been provided in this follow-up work. Finally, it will be interesting to explore the connection between our counting formula with those proposed in earlier works \cite{luiz,jelena1,jelena2,jelena3}.

%%%%%%%%%%%%%%%%%%%%%%%%%%%%

\paragraph{\textbf{Acknowledgments --}}

We thank Atri Bhattacharya, Sauri Bhattacharyya, Fiona Burnell, Chen-Hsuan Hsu, and Sayeh Rajabi for stimulating discussions. We are also grateful
to Sudip Chakravarty, Pinaki Majumdar, and Arijit Saha for their valuable comments on the
manuscript. This research was partially supported by the Templeton Foundation.
Research at the Perimeter Institute is supported
in part by the Government of Canada
through Industry Canada, and by the Province of Ontario through the
Ministry of Research and Information.

%%%%%%%%%%%%%%%%%%%%%%%%%%%%%%%%5

\bibliography{ref}

%%%%%%%%%%%%

%%\begin{widetext}
%\appendix*{Representation of the Clifford Algebra}
%\label{app:clifford}
%
%In this appendix, we summarize the representations of the Clifford algebra in terms of the $\gamma$-matrices, which have been used to write down the generic BdG Hamiltonian in the BDI class in Sec.~\ref{pointb}. Let us define the $\gamma$-matrices for $d=2 \ell$ (even) as follows:
%\begin{eqnarray}
%\gamma^{2j -1 } &=& \underbrace{\mathbb{I} \otimes \cdots \mathbb{I}}_{j-1}
%\otimes  \sigma^1 \otimes 
%\underbrace{\sigma^3 \otimes \cdots \otimes \sigma^3}_{ \ell-j} \,,\nn
%%%%%%%%%%%
%\gamma^{2j } &=& \underbrace{\mathbb{I} \otimes \cdots \mathbb{I}}_{j-1}
%\otimes  \sigma^2 \otimes 
%\underbrace{\sigma^3 \otimes \cdots \otimes \sigma^3}_{  \ell-j} \,,\nn
%\end{eqnarray}
%%%%%%%%%%%%%%%%%%%%%
%where $j=1, \ldots,  \ell $.
%If $d=2 \ell+1$ (odd), then we need to add another $\gamma$-matrix
%\begin{eqnarray}
%\gamma^{2 \ell+1 } &=& \underbrace{\sigma^3 \otimes \cdots \sigma^3}_{ \ell}
%=(- i )^n \gamma^1 \cdots \gamma^{2 \ell} \,.
%\end{eqnarray}
%%%%%%%%%%%%%%%%
%Defining
%\begin{eqnarray}
%\Gamma = \gamma^1 \gamma^3 \cdots \gamma^{2 \ell+1} \,,
%\label{Gamma}
%\end{eqnarray}
%%%%%%%%%%%%%%%
%we get
%\begin{eqnarray}
%\Gamma^2 &=& (-1)^{ \ell \,( \ell+1)/2} \,,\nn
%\left( \gamma^{\mu} \right)^T &=& \left( \gamma^{\mu} \right)^{*} =(-1)^ \ell \Gamma \gamma^\mu \Gamma^{-1} \,,
%\end{eqnarray}
%where $\mu = 1,2 , \ldots , 2 \ell , 2 \ell+1 $.
%

\end{document}